\documentclass[aps,twocolumn,floatfix,showpacs]{revtex4}
\usepackage{amsmath,amssymb,graphicx}
\begin{document}
 
\title{Incompleteness of  the Thouless, Anderson,
and Palmer mean-field description of the spin-glass phase}

\author{V.  Jani\v{s}}

\affiliation{Institute of Physics, Academy of Sciences of the Czech
  Republic, Na Slovance 2, CZ-18221 Praha 8, Czech Republic}

\date{\today}

\begin{abstract}
We analyze the low-temperature behavior of mean-field equations of
Thouless, Anderson, and Palmer (TAP). We demonstrate that degeneracy in
free energy makes  the low-temperature TAP states unstable. Different
solutions of the TAP equations, independent in the TAP approach, become
coupled if an infinitesimal interaction between them is introduced. By
means of real spin replicas we derive a self-averaging free energy free of
unstable states with local magnetizations and homogeneous overlap
susceptibilities between different spin replicas as order parameters. We
thereby extend the TAP approach to a consistent description of the
spin-glass phase for all configurations of spin exchange with
(marginally) stable and thermodynamically homogeneous free energy.
\end{abstract}
\pacs{64.60.Cn,75.50.Lk}
\maketitle

\section{Introduction}\label{sec:Intro}

The Parisi replica-symmetry breaking (RSB) scheme \cite{Parisi80} was
proved to be an exact solution of the Sherrington-Kirkpatrick (SK) model
of spin glasses \cite{Talagrand04}.  The analytic form of the mean-field
theory of Ising spin glasses is hence known. What has not yet  been
unambiguously identified is the physical origin of the order parameters
from the RSB solution of the  replica trick. The replica trick is used to
allow averaging of  free energy over random configurations of spin
couplings. Thermal and disorder-induced fluctuations are summed in the
replica trick simultaneously via a single averaging of an $n$-times
replicated partition function. One is hence unable to determine whether
the former or the latter fluctuations give rise to the order parameters
from the Parisi solution. To find the physical origin of the order
parameters of the RSB solution one must separate the thermal and the
disorder-induced fluctuations.
 
The direct thermodynamic approach summing separately the thermal
fluctuations for fixed typical configurations of spin couplings $J_{ij}$ in
the SK model was pioneered by Thouless, Anderson, and Palmer  \cite{TAP77}.
The standard TAP theory of the SK model contains only local magnetizations
$m_i$  as order parameters. The averaging of the TAP free energy over
random configurations within linear-response theory and with the
fluctuation-dissipation theorem leads to the (replica-symmetric) SK
solution unstable in the low-temperature phase \cite{Sommers78}. That is,
no Parisi RSB parameters emerge directly in the TAP theory.

The assumptions made for the averaging over randomness in the TAP theory
are essentially equivalent to uniqueness of the equilibrium state for each
relevant configuration of spin couplings. It appeared rather soon, however,
that the TAP equations display a multitude of solutions in the spin-glass
phase \cite{Bray79} resulting in a complex free-energy landscape of
quasi-equilibrium states \cite{Mezard87}. The existence of multiple
solutions of the TAP equations would not pose a problem if different
states were distinguishable by symmetry-breaking fields introduced in
free energy. The solutions of the TAP equations in the spin-glass phase
are highly degenerate in free energy and cannot be singled out by external
fields. Even worse is the fact that for a large number of configurations of
spin couplings there are no stable states, local minima of the TAP free
energy \cite{Bray79,Nemoto85}. One hence cannot define a unique
macroscopic thermodynamically stable state for these configurations. The
existence of an exponentially large number of solutions of the mean-field
equations has become a hallmark of spin-glass models. A new branch of
research on complexity of solutions in the mean-field theory of spin
glasses  emerged \cite{Bray80,Aspelmeier04,Cavagna04,Crisanti04}.

The nonexistence of thermodynamically stable macroscopic states for
majority of  configurations of  spin couplings hinders the
existence of the thermodynamic limit in the TAP approach. To circumvent
this problem De Dominicis and Young suggested that the equilibrium state
in the TAP approach be defined as a weighted sum over different TAP
solutions \cite{DeDominicis83} That is, one assumes that the partition
function can be represented as
\begin{equation}\label{eq:TAP-EQ}
\text{Tr}_S\exp\left[-\beta H\{S\}\right] = \sum_\alpha^\mathcal{N}
\exp\left[-\beta F_{TAP}\{m_i^\alpha\}\right]\ ,
\end{equation}
where $\mathcal{N}$ is the number of TAP solutions labeled by superscript
$\alpha$. Assumption~\eqref{eq:TAP-EQ} means that the phase space of the
SK model is effectively disconnected.  It consists of pockets of spin
configurations corresponding to different TAP solutions separated by
impenetrable infinite energy barriers.

Albeit assumption~\eqref{eq:TAP-EQ} defines a relation between
individual TAP solutions and the macroscopic thermodynamic state, it
does not introduce the RSB order parameters. They emerge in the De
Dominicis and Young completion of the TAP theory when the replica
trick for averaging over random configurations of spin couplings is
used. Without averaging over randomness we are able neither to verify
Eq.~\eqref{eq:TAP-EQ} nor to trace down the genesis of the RSB order
parameters beyond the replica trick.

Averaging over randomness should not generally be the  eventual tool for
introducing  the RSB order parameters.  Guerra and Toninelli recently
proved that the free energy of the SK model is self-averaging
\cite{Guerra02}. Should the TAP approach be exact, one had to trace down
the Parisi order parameters within the TAP approach without resorting to
averaging over randomness. A question then arises whether the TAP
construction indeed provides a complete  description of the thermodynamics
of the SK model.
 
We know that to derive the TAP theory we have to assume uniqueness of
the thermodynamic equilibrium state described by a set of local
magnetizations. This, however, is the case only if a convergence
condition for the linked-cluster expansion
\begin{equation}\label{eq:AT-normal}
1  \ge  \frac {\beta^2J^2}N \sum_i  (1 - m_i^2)^2
\end{equation}
holds \cite{Plefka82}. Equality in the above condition determines the de
Almeida-Thouless (AT) line separating the high-temperature from the
spin-glass phase along which the spin-glass susceptibility diverges
\cite{AT78}. Condition~\eqref{eq:AT-normal} is broken below the AT line
for a macroscopic portion of spin-coupling configurations and the TAP free
energy does not have an adequate (rigorous) justification there. We must
continue analytically the TAP thermodynamic potentials from the
high-temperature phase, where Eq.~\eqref{eq:AT-normal} is obeyed, to the
low-temperature one, where the latter condition may be broken. Such a
procedure is not uniquely defined, unless we  have appropriate
symmetry-breaking fields at our disposal. Presently, it is assumed that
there are only local magnetic fields, Legendre conjugates to the local
magnetizations, as symmetry-breaking forces. The TAP free energy in the
spin-glass phase  consequently has the same form as in the
high-temperature phase, i.~e., it is described by the same order
parameters, local magnetizations $m_i$.

Recently Plefka suggested that the TAP equations in situations with
unstable states where Eq.~\eqref{eq:AT-normal} is broken should be
stabilized by introducing a new "order parameter", a correction to the
local magnetic susceptibility beyond the fluctuation-dissipation theorem
\cite{Plefka02a}. Plefka's extended solution, however, does not allow for
a diagrammatic representation, the order parameter for the deviation from
the fluctuation-dissipation theorem cannot be derived from free energy,
and hence a physical meaning cannot be given to the calculations
containing the TAP solutions breaking condition~\eqref{eq:AT-normal}.
Although the unstable states seem to become marginally stable in the
thermodynamic limit \cite{Plefka02b}, the number of states breaking
condition~\eqref{eq:AT-normal} linearly increases with the number of
lattice sites and diverges in the thermodynamic limit
\cite{Bray79,Nemoto85}. Unstable states from large but finite volumes
hence remain statistically relevant also in the thermodynamic limit, since
the negative values of the r.h.s. of Eq.~\eqref{eq:AT-normal}vanish with
power $N^{-2/3}$ \cite{Plefka02b,Cavagna04}. We hence cannot disregard or
inappropriately treat the finite-volume unstable states without further
considerations. We can deduce that the number of TAP configurations with
unstable states is macroscopically relevant in the thermodynamic limit
also indirectly when averaging  the TAP free energy over spin couplings
$J_{ij}$. Using linear response and the fluctuation-dissipation theorem,
equivalent to self-averaging property of free energy of ergodic systems,
we fail to produce a thermodynamically stable equilibrium state in the
spin-glass phase. Since we know that the exact free energy of the SK model
is self-averaging, the TAP construction breaks down in the spin-glass
phase. To attain a self-averaging configurationally-dependent free energy
we must extend consistently the TAP free energy also to configurations
with unstable states, i.~e., beyond the validity of
inequality~\eqref{eq:AT-normal}.
 
The aim of this paper is to  demonstrate that the TAP free energy becomes
unstable whenever stability condition~\eqref{eq:AT-normal} is broken and
the TAP equations do not have a single solution independent of the initial
conditions.   By using spin replicas for portions of the phase space
belonging to different TAP solutions we show that linear response theory
is broken when an infinitesimal interaction between different spin replicas
(solutions of the TAP equations) is introduced. This breakdown generates a
set of new homogeneous order parameters, overlap susceptibilities between
different replicas.  They lift degeneracy in the TAP free energy and break
independence of different solutions of the TAP equations. We derive
a generalization of the TAP free energy for one configuration of spin
couplings containing site-dependent local magnetizations $M_i$ and
homogeneous local overlap susceptibilities $\chi^{ab}$ as order
parameters. The latter are directly related to the RSB order parameters of
the Parisi solution. In the paramagnetic phase $\chi^{ab} = 0$ and we
recover the TAP free energy.  In the low-temperature phase, for
configurations of spin couplings for which condition~\eqref{eq:AT-normal}
is broken, the overlap susceptibilities become nonzero and we observe
macroscopic deviations from the TAP free energy. Different solutions of
the TAP equations are hence  not separated by infinite energy barriers.
Mutual thermodynamically induced interaction between solutions of the TAP
equations mediated by the overlap susceptibilities interconnects  parts of
the phase space separated in the TAP theory.   The phase space becomes
simply connected and  stable macroscopic thermodynamic states exist
for each configuration of spin couplings independently of whether
condition~\eqref{eq:AT-normal} is fulfilled or not.  The interaction
between different TAP states also leads to the existence of a single
equilibrium state with a well defined  thermodynamic limit  generated from
 a self-averaging free energy functional.
 
 The paper is organized as follows. In Sec.~\ref {sec:TAP-standard} we
 recall the basic ingredients of the TAP theory with restrictions on
 its applicability. We use real replicas and the demand of
 thermodynamic homogeneity to extend (analytically continue) the TAP
 approach to situations with unstable TAP states in
 Sec.~\ref{sec:Replicated-TAP}. In Sec.~\ref{sec:RS-TAP} we reduce the
 general theory to one hierarchical level and present the modified TAP
 equations, study their stability and finally demonstrate explicitly
 near the critical point that the TAP construction indeed becomes
 unstable in the spin-glass phase. In the last section we summarize
 our findings and discuss their consequences.

\section{TAP mean-field theory and stability of its equilibrium states}
\label{sec:TAP-standard}

We first recall the basic  concepts of  the  TAP theory for the SK model so
that we  understand the restrictions under which the TAP theory is
applicable. In the diagrammatic representation the TAP free energy was
derived as a sum of  tree and single-loop (cavity-field) contributions
with specific restrictions of the SK model on spin couplings $J_{ij}$,
namely $\sum_j J_{ij}^{2n + 1} = 0$ and $\sum_j J_{ij}^2 =
J^2$~\cite{Janis05b}. Due to the fluctuation-dissipation theorem the local
susceptibility containing the loop contributions is a function of the local
magnetization and  the TAP free energy for the SK model is a functional of
only local magnetizations $m_i$. It is convenient to represent the TAP
free energy in the following form
\begin{multline}\label{eq:TAP-FE}
F_{TAP} = \sum_i\left\{ m_i\eta_i^0  -
\frac 1\beta  \ln 2 \cosh[\beta(h + \eta_i^0)]\right\}   \\  %
- \frac 12 \sum_{ij}\left[J_{ij} m_i m_j + \frac 12 \beta
J_{ij}^2 (1 - m_i^2)(1 - m_j^2)\right]
\end{multline}
where we introduced apart from local magnetizations $m_i$ also
internal inhomogeneous magnetic field $\eta_i^0$. The sets of
parameters $m_i$ and $\eta_i^0$ are Legendre conjugate variables and
are treated variationally in free energy~\eqref{eq:TAP-FE}. That is,
they have to determine an extremal value of this free-energy
functional.  The corresponding stationarity (TAP) equations for these
parameters read
\begin{subequations}\label{eq:TAP-equations}
\begin{align}
\label{eq:TAP-magnetization}
m_i &= \tanh[\beta (h + \eta_i^0)]\ , \\
\label{eq:TAP-magf}
\eta_i^0 &= \sum_j J_{ij} m_j - m_i  \sum_j \beta J_{ij}^2(1 - m_j^2)\ .
\end{align}
\end{subequations}
These equations can now be solved numerically for finite numbers of lattice
sites and given configurations of spin couplings. But not all solutions of
equations~\eqref{eq:TAP-equations} are physical ones. Only locally stable
solutions for which the nonlocal susceptibility does not contain negative
eigenvalues are meaningful.  The inverse of the susceptibility is defined
as a second derivative of free energy~\eqref{eq:TAP-FE}
%
\begin{multline}\label{eq:TAP-isuscept}
  \left(\chi^{-1}\right)_{ij} = \frac {\partial^2 \beta F_{TAP}}
  {\partial m_i \partial m_j} + \sum_l \left[\frac{\partial^2\beta
      F_{TAP}} {\partial m_i\partial \eta_l^0} \frac
    {\partial\eta_l^0}{\partial m_j}\right. \\ \left. +
    \frac{\partial^2\beta F_{TAP}} {\partial m_j\partial \eta_l^0}
    \frac {\partial\eta_l^0} {\partial m_i}\right] + \sum_{kl} \frac
  {\partial^2 \beta F_{TAP}}{\partial \eta_k^0 \partial \eta_l^0}
  \frac {\partial \eta_k^0}{\partial m_i} \frac {\partial \eta_l^0}
  {\partial m_j} \\ = - \beta J_{ij} + \delta_{ij}\left(\frac 1{1 -
      m_i^2} + \sum_l \beta^2 J_{il}^2 (1 - m_l^2)\right)\ .
\end{multline}
That is, only local minima of the TAP free energy~\eqref{eq:TAP-FE} as a
functional of local magnetizations $m_i$, when the internal magnetic
fields are resolved, are physically acceptable.

Non-negativity of the eigenvalues of the linear susceptibility is not the
only stability criterion.  There is  a stronger condition on consistency
of the TAP theory. It is connected with the existence of a non-degenerate
equilibrium state, an assumption used in the derivation of the TAP free
energy. This condition is expressed as positivity of the spin-glass
susceptibility $\chi_{SG}$. It is easy to find by summing the
leading-order ($N^{-1}$) diagrammatic contributions \cite{Bray79}  that the
spin-glass susceptibility has in the SK model the following representation
\begin{subequations}
\begin{equation}\label{eq:TAP-chi2}
\chi_{SG} \equiv \frac 1N\sum_{ij}\chi_{ij}^2 = \frac 1N \sum_i\frac
{\chi_{ii}^2} {1 - \sum_j \beta^2 J_{ij}^2 \chi_{jj}^2}\ .
\end{equation}
This representation of the spin-glass susceptibility was derived
diagrammatically but it  is valid quite generally as long as the r.h.s of
Eq.~\eqref{eq:TAP-chi2} remains nonnegative, that is if
\begin{equation}\label{eq:TAP-chi2-stabil}
1 \ge \sum_j \beta^2 J_{ij}^2 \chi_{jj}^2\ .
\end{equation}
We show in Appendix~\ref{sec:App-resolvent} that
representation~\eqref{eq:TAP-chi2} can be derived also non-perturbatively
using a theorem of Pastur and continuity of the resolvent for the inverse
nonlocal susceptibility.

Realizing that the local susceptibility in the TAP theory reads
\begin{equation}\label{eq:TAP-chi-loc}
\chi_{ii} = 1 - m_i^2
\end{equation}
\end{subequations}
we find that the stability condition from Eq.~\eqref{eq:AT-normal} equals
the condition on positivity of the spin-glass susceptibility,
Eq.~\eqref{eq:TAP-chi2-stabil}. Positivity of the spin-glass susceptibility is
a feature that each consistent solution must possess. If it is broken, then
the phase space of the order parameters is incomplete and some relevant
fluctuations have not been taken into account appropriately. Note that
in general positivity of the spin-glass susceptibility does not coincide
with positivity of the eigenvalues of the nonlocal susceptibility. Only
squares of the eigenvalues of the latter contribute to the former.
The spin-glass susceptibility may become negative even if the linear
susceptibility is positive, that is for a local minimum of the TAP free
energy.
 
The TAP theory was derived assuming that the resulting free energy
leads to a single (non-degenerate) stable thermodynamic state. That
is, the TAP equations~\eqref{eq:TAP-equations} lead to a single
physical solution that can be separated from nonphysical ones by
finite energy gaps. We know, however, that this is not the case in the
spin-glass phase.  Hence the TAP free energy is internally consistent
only in the high-temperature phase, where it leads to a single stable
equilibrium state. One has to be more careful when extending the TAP
approach to the low-temperature phase. There we cannot separate the
physical solutions of the TAP equations from the nonphysical ones
breaking stability condition~\eqref{eq:AT-normal}. We have to modify
the TAP approach to situations with many quasi-equilibrium and
unstable states degenerate in free energy.

\section{Thermodynamic homogeneity and multiple TAP states}
\label{sec:Replicated-TAP}

The existence of many solutions of the TAP equations degenerate in
free energy hinders the existence of a stable macroscopic equilibrium
state and does not allow to perform the thermodynamic limit. In a
degenerate case we cannot fix a single solution when enlarging the
volume of the system and large fluctuation do not extinguish in the
thermodynamic limit. Different unstable solutions of the TAP equations
degenerate in free energy can be distinguished only by initial
conditions, being the only input to Eqs.~\eqref{eq:TAP-equations}.
This means that the TAP free energy is effectively not
thermodynamically homogeneous, since it does not depend only on
spatial densities of extensive variables.  One way to handle
a multitude of  quasi-equilibrium states in the TAP approach is to
assume infinite barriers between different TAP states (independence of
different solutions of the TAP equations) and use
Eq.~\eqref{eq:TAP-EQ}. We can, however, avoid
assumption~\eqref{eq:TAP-EQ} in that we do not a priori exclude
interaction between different TAP states. Since different solutions of
the TAP equations belong in the beginning to independent separate
parts of the phase space, we can introduce for each TAP solution its
own replica of the spin variables and sum up thermal fluctuations for
each solution separately. This is actually the concept of real
replicas that has been used by the author to derive the RSB solution
from the demand of thermodynamic homogeneity of the averaged free
energy \cite{Janis05a}. In the TAP approach without averaging over
randomness we can give a transparent physical interpretation to real
spin replicas.

Let us assume that we have $\nu$ different TAP solutions
(distinguished by their history). Since different solutions are
initially thermodynamically independent we introduce independent spin
replica for each TAP solution and replicate $\nu$-times the original
phase space. The partition function on this replicated phase space can
be represented as $\left[\text{Tr}\ e^{-\beta H}\right]^\nu =
\text{Tr}_\nu \exp\left\{\beta\sum\limits_{a=1}^{\nu}H^\alpha\right\}
= \text{Tr}_\nu \exp\left\{\beta\sum\limits_{a=1}^{\nu}
  \left(\sum_{i,j} J_{ij}S_i^a S_j^a + \sum_i S_i^a\right)\right\}$,
where each replicated spin variable $S_i^a$ is treated independently,
i.~e., the trace operator $\text{Tr}_\nu$ operates on the $\nu$-times
replicated phase space. The free energy of an $\nu$-times replicated
system is just $\nu$-times the free energy of the non-replicated one,
if it is thermodynamically homogeneous. We now break independence of
individual spin replicas and add a small (infinitesimal) homogeneous
perturbation breaking the replica independence $\Delta H(\mu)= \sum_i
\sum_{a < b} \mu^{ab} S_i^a S_i^b$. We could also break the replica
independence inhomogeneously by a site-dependent symmetry-breaking
field $\mu_{ii}^{ab}$. Since the stability condition for the TAP
theory, Eq.~\eqref{eq:AT-normal}, is global, we are effectively able
to break the replica dependence only globally as we demonstrate in the
next section.
 
It is not the field $\mu^{ab}$ connecting different replicas that is of
physical interest. We are interested  in the linear response of the
system to this perturbation. We derived \cite{Janis05b} that after
switching off the field $\mu^{ab}$ the $\nu$-times replicated TAP free
energy reads
\begin{widetext}
\begin{multline}\label{eq:nuTAP}  F_\nu =\frac 1\nu
\sum_{a=1}^\nu\left\{\sum_i M_i^a\left[\eta_i^a + \beta J^2
\sum_{b=1}^{a-1} \chi^{ab}M_i^b\right] - \frac 14\sum_{i,j} \beta
J_{ij}^2\left[1 - (M_i^a)^2\right]\left[1-(M_j^a)^2 \right] - \frac
12\sum_{i,j} J_{ij}M_i^a M_j^a \right. \\ \left. + \frac{\beta J^2N}2
\sum_{b=1}^{a-1} (\chi^{ab})^2 \right\} - \frac 1{\beta \nu} \sum_i \ln
\mbox{Tr}\exp\left\{ \beta^2 J^2\sum_{a < b}^\nu \chi^{ab}S_i^aS_i^b
+\beta\sum_{a=1}^\nu\left(h+\eta_i^a\right) S_i^a\right\}\ .
  \end{multline} \end{widetext}
In this expression local magnetizations $M^a_i$ and local internal magnetic
fields $\eta^a_i$ are configurationally dependent Legendre conjugate
variational variables determined from stationarity equations analogously
to the TAP equations~\eqref{eq:TAP-equations}. Apart from these parameters
we introduced $\chi^{ab}, a\neq b$, averaged overlap local
susceptibilities representing a linear response to the replica-mixing
field $\mu^{ab}$. They are global (translationally invariant) variational
variables, Legendre conjugates to the symmetry breaking fields $\mu^{ab}$.
It is straightforward to verify that at the saddle point we have
$\chi^{ab}= N^{-1}\sum_i\left[\langle S^a_iS^b_i\rangle_T - \langle
S^a_i\rangle_T\langle S^b_i\rangle_T\right]$, where
$\langle\ldots\rangle_T$ stands for thermal averaging.

Free energy $F_\nu$ from Eq.~\eqref{eq:nuTAP} becomes independent of
the replication index $\nu$ and reduces to the TAP free energy if
$\chi^{ab}=0$. This is just the case when the convergence criterion
for the TAP theory, Eq.~\eqref{eq:AT-normal}, holds. A difference
between the original TAP free energy and that from
Eq.~\eqref{eq:nuTAP} emerges only in regions with unstable states in
the TAP equations. Free energy~\eqref{eq:nuTAP} can hence be viewed
upon as a general form of the TAP-like free energy for one
configuration of spin couplings. Different replica indices correspond
to different solutions of mean-field equations. Unlike the TAP
approach the different states in free energy~\eqref{eq:nuTAP} are
allowed to interact via the overlap susceptibility $\chi^{ab}$.

 If free energy $F_\nu$ is thermodynamically homogeneous it should not
dependent on the replication parameter $\nu$. We already know that this
is not the case, at least for the averaged TAP free energy,  when stability
condition~\eqref{eq:AT-normal} is broken \cite{Janis05a}. If thermodynamic
homogeneity is broken we have to use the new order parameters so as to
restore this fundamental property. Only thermally homogeneous systems
possess non-degenerate stable equilibrium states extremizing a free-energy
functional and can be extended uniquely to infinite volumes. In our
construction, it is the matrix of overlap susceptibilities that should
restore thermodynamic homogeneity in the TAP approach.

We now impose the condition of thermodynamic homogeneity on free
energy~\eqref{eq:nuTAP} in that we demand the existence of a unique
thermodynamic state. That is, all spin replicas must be equivalent and
must lead to the same order parameters. This property can be quantified as
follows
\begin{subequations}\label{eq:equivalence-replicas}
\begin{align}\label{eq:equivalence-magnetization}
M_i^a \equiv \langle S_i^a\rangle_T & = M_i\ ,  \\
\label{eq:equivalence:symmetry} \chi^{ab} & = \chi^{ba}\ , \\
\label{eq:equivalence-different} \{\chi^{a1},\ldots, \chi^{a\nu}\} & =
\{\chi^{b1},\ldots, \chi^{b\nu}\}\ .
\end{align}\end{subequations}
Equation~\eqref{eq:equivalence-magnetization} says that at the level of
local magnetizations different spin replicas are indistinguishable. That
is, the internal local magnetic  fields are replica independent, $\eta_i^a
=\eta_i$. Conditions~\eqref{eq:equivalence:symmetry}
and~\eqref{eq:equivalence-different} restrict  the matrix of overlap
susceptibilities to be symmetric with  rows (columns) being only
permutations of each other. We remind that $\chi^{aa} = 0$. The matrix
$\chi^{ab}$ contains then only $\nu -1$ independent parameters. that can be
cast into groups of identical values. If we set $\nu_K > \nu_{K-1}
>\ldots \nu_! > 1$ we may choose  $\nu_1 - 1$-times a value $\chi_1$,
$(\nu_{2} - \nu_{1})$-times an overlap $\chi_{2}$,  and so on up to.
$(\nu_K - \nu_{K-1})$-times an overlap $\chi_K$.

As the last step we have to determine the structure of the matrix
$\chi^{ab}$ with the above restrictions that would lead to an analytic
free-energy functional of variables $\nu_1,\ldots, \nu_K$ and
$\chi_1,\ldots, \chi_K$. The easiest way to determine the most general
available structure of $\chi^{ab}$ is to use a hierarchical construction.
It starts with $K=1$ and increases the number of different values of the
overlap susceptibilities only if the solution with $K$ different values
becomes unstable. In the case $K=1$ the matrix of the overlap
susceptibilities is uniquely determined by a multiplicity $\nu_1$ of the
only value $\chi_1$. We examine this particular case in detail in the
next section. If the theory with $K=1$ is unstable, we build up a theory
with $K=2$ values of the overlap susceptibility, $\chi_1$ and $\chi_2$.
We assume that not only the individual replicas are equivalent but also
blocks of replicas describing the solution with $K=1$ are  equivalent.
That is, the diagonal elements in the solution with $K=1$ are replaced by
matrices $\nu_1\times\nu_1$ with zero on the diagonal and $\chi_1$ on the
off-diagonal positions.  The remaining  off-diagonal elements in the
solution with $K=2$ are filled with the value $\chi_2$. In this way we go
on to higher hierarchies. We end up with an ultrametric structure of the
Parisi RSB solution.  It is of essential importance that the ultrametric
structure allows for  an analytic representation of the hierarchical free
energy with $K$ different values of the overlap susceptibility. In fact,
the ultrametric arrangement of the overlap susceptibilities $\chi^{ab}$
seems to be the  most general structure in which the free energy is an
analytic function of parameters $\chi_l, \nu_l$ for $l = 1,\ldots, K$.

Inserting the ultrametric structure with $K$ hierarchies of
$\chi^{ab}$ in Eq.~\eqref{eq:nuTAP} and after $K$-times applied the
Hubbard-Stratonovich transformation linearizing the spin variables in
the exponent of $\exp\{\beta^2J^2\sum_{a<b}\chi^{ab}S_i^aS_i^b\}$ we
obtain an analytic representation of the $K$-level hierarchical
generalization of the TAP free energy
\begin{widetext}
\begin{multline}\label{eq:RSB-TAP}
F_K(\chi_1,\nu_1,\ldots,\chi_K,\nu_K) = - \frac 14\sum_{i,j} \beta
J_{ij}^2(1 - M_i^2)(1 - M_j^2) - \frac 12\sum_{i,j} J_{ij}M_i M_j \\ +
\sum_i M_i \left[\eta_i + \frac 12 \beta J^2 M_i \sum_{l=1}^K (\nu_l -
\nu_{l-1})\chi_l\right] + \frac{\beta J^2N}4 \sum_{l=1}^K(\nu_l -
\nu_{l-1})\chi_l^2 + \frac{\beta J^2N}2 \chi_1  \\ - \frac
1{\beta\nu_K}\sum_i \ln\left[\int_{-\infty}^{\infty}\mathcal{D}\lambda_K
\left\{\dots \int_{-\infty}^{\infty} \mathcal{D}\lambda_1
\left\{2\cosh\left[\beta\left(h + \eta_i + \sum_{l=1}^{K}\lambda_l
\sqrt{\chi_l - \chi_{l+1}}\right)\right]\right\}^{\nu_1}\ldots\right\}
^{\nu_K/\nu_{K-1}}\right]\ .
\end{multline}
\end{widetext}
 We  abbreviated  $\mathcal{D}\lambda_l \equiv {\rm d}\lambda_l\
e^{-\lambda_l^2/2}/ \sqrt{2\pi}$ and used $\nu_{0} = 1, \chi_{K+1} =0$.
Notice that in  our derivation  $\nu_1 < \nu_2 <\ldots <\nu_K = \nu$ and
$\chi_1 > \chi_2 >\ldots >\chi_K\ge 0$.  Free energy~\eqref{eq:RSB-TAP}
should be an extremum with respect to  matrix $\chi^{ab}$ so that a
thermodynamically homogeneous free energy is produced. Thermodynamic
homogeneity is achieved in  free energy~\eqref{eq:RSB-TAP} if it does not
depend on $\nu_K$. This is equivalent to vanishing of $\chi_K$. Since the
trivial solution $\chi_l = 0$ always satisfies the stationarity equations
for any $l=1,\ldots,K$, free energy~\eqref{eq:RSB-TAP}  with $K$
hierarchies is thermodynamically homogeneous if $\chi_K =0$ is the only
physical solution of the respective stationarity equation. Nonexistence of
a nontrivial solution for $\chi_K$ determines the number of hierarchical
levels needed to achieve a globally stable solution.

Both sets of parameters $\chi_l$ and $\nu_l$ must be treated
variationally and their physical values must be determined from
respective stationarity equations. The equilibrium multiplicity
factors $\nu_l^{eq}$, determined from  $\partial
F_K/\partial \nu_l = 0$, no longer need be integers, form an
increasing sequence, and they even can be smaller than one. As
discussed in Ref.~\onlinecite{Janis05a} the stationarity equations for
$\nu_l$ have two solutions, $\nu_l^{eq} \ge 1$ and $\nu_l^{eq} \le 1$.
The latter case is actually the physical one, since it minimizes
thermodynamic inhomogeneity, if occurs. The value $\nu_l < 1$
determines then a portion of the phase space (relative number of lattice
sites) of one TAP solution influenced by the existence of other TAP
solutions. With a homogeneous, site-independent overlap susceptibility
all spins in each solution are equivalent. The exponent $\nu_l$ then
says that $\nu N$ spins on average are influenced by other TAP
solutions \cite{Janis05a}.

Free energy~\eqref{eq:RSB-TAP} is the most general analytic
continuation of the TAP free energy to the low-temperature phase. If
condition~\eqref{eq:AT-normal} is obeyed for $\chi_l =0, l = 1,\ldots,
K$ and $F_K(\chi_1,\nu_1,\ldots,\chi_K,\nu_K) = F_{TAP}$. Free energy
$F_K$ is self-averaging and it is numerically identical with the RSB
free energy with $K$ hierarchical levels as derived in
Ref.~\cite{Janis05a}. In the extension of the TAP theory,
Eq.~\eqref{eq:RSB-TAP}, the RSB order parameters are induced by
thermal fluctuations and serve as mediators of interaction between
different TAP solutions.

\section{One-level hierarchical TAP theory}
\label{sec:RS-TAP}

Representation~\eqref{eq:RSB-TAP} of a configurationally dependent
free energy is rather complicated. It is a futile activity to try to
solve the corresponding stationarity equations for a chosen
configuration of spin couplings in full generality before exploring
suitable simplifications.  Moreover, it is not necessary to
reconstruct the complete spatial distributions of site-dependent local
magnetizations when we are interested in thermodynamic quantities
determined by only lattice sums. Since free
energy~\eqref{eq:RSB-TAP} is self-averaging, in most situations we can
replace the sums over the lattice sites by averages over the
distribution of random spin couplings. Thereby we perform this
averaging within linear response theory and with the
fluctuation-dissipation theorem. That is, we use the same averaging
rules to Eq.~\eqref{eq:RSB-TAP} as used on $F_{TAP}$ in deriving the
SK solution. This direct way of averaging of $F_K$ leads to the Parisi
solution with $K$ hierarchical levels \cite{Janis05b,Janis05a}.

To demonstrate explicitly that free energy~\eqref{eq:RSB-TAP} is a
nontrivial extension of the TAP free energy in the low-temperature
phase also for fixed configurations of spin couplings we resort our
analysis of this free energy to the solution with $K=1$, that is, to
the one-level hierarchical solution.

\begin{widetext}
\subsection{Stationarity equations}
\label{sec:RS-TAP-EQ}

It is straightforward to reduce the general expression for the hierarchical
free energy $F_K$ to the case $K=1$ with $\chi_1 = \chi$ and $\nu_1 = \nu$.
We obtain
\begin{multline}\label{:eq:1RSB-TAP}
 F_1(\chi,\nu) = - \frac 14\sum_{i,j} \beta J_{ij}^2(1 - M_i^2)(1 - M_j^2)
- \frac 12\sum_{i,j} J_{ij}M_i  M_j  + \frac{\beta J^2N}4 \chi[(\nu -
1)\chi + 2] \\ + \sum_i M_i \left[\eta_i + \frac 12 \beta J^2(\nu - 1)
\chi M_i \right]  - \frac 1{\beta \nu} \sum_i \ln\int \mathcal{D}\lambda_i
\big[2\cosh [\beta(h + \lambda_i J\sqrt{\chi} + \eta_i) ]\big]^\nu \ .
 \end{multline}  \end{widetext}
Free energy $F_1(\chi,\nu)$  is represented in
closed form and is analytic in all its variables $M_i$, $\eta_i$, $\chi$,
and $\nu$. It reduces to the TAP expression if $\chi =0$, which is the
case when Eq~\eqref{eq:AT-normal} is fulfilled by the local magnetizations
$M_i$.

The stationarity equation for the site-dependent local magnetization
follows from $\partial F_1/\partial \eta_i = 0$ from which we obtain
\begin{subequations}\label{eq:Local-equations}
 \begin{align}\label{eq:Local-mag}
   M_i \nonumber  & = \left\langle \rho^{(\nu)}(h + \eta_i;\lambda,\chi )
\tanh[\beta(h + \eta_i + \lambda J\sqrt{\chi})] \right\rangle_\lambda
\nonumber \\
& \equiv  \langle \rho_i^{(\nu)} t_i\rangle_\lambda\ ,
 \end{align}
 where
\begin{multline}\label{eq:Density-def}
 \rho_i^  \nu \equiv \rho^{(\nu)}(h + \eta_i ; \lambda,\chi) \\ =
\frac{\cosh^\nu[\beta(h + \eta_i + \lambda J\sqrt{\chi})]} {\left\langle
\cosh^\nu[\beta(h + \eta_i + \lambda J\sqrt{\chi} )]\right\rangle_\lambda}
\end{multline}
is a density matrix.  We denoted $\langle
X(\lambda)\rangle_\lambda = \int \mathcal{D}\lambda X(\lambda)$.

The internal local magnetic field $\eta_i$ is determined from
$\partial F_1/\partial M_i = 0$ which results in
\begin{equation}\label{eq:Local-eta}
\eta_i = \sum_jJ_{ij}M_j - M_i\left[\beta J^2(\nu - 1)\chi + \sum_j \beta
J_{ij}^2 (1 - M_j^2)\right] \ .
\end{equation}\end{subequations}

In addition to the site-dependent order parameters we have to determine the
physical (stationary) values of the homogeneous parameters $\chi$ and
$\nu$. From
the equation $\partial F_1/\partial \chi = 0$ we obtain
\begin{subequations}\label{eq:Global-equations}
\begin{equation}\label{eq:Global-chi}
\chi = \frac 1N\sum_i \left[\left\langle\rho_i^{(\nu)}
t_i^2\right\rangle_\lambda - \left\langle\rho_i^{(\nu)}
t_i\right\rangle_\lambda^2\right] \ .
\end{equation}

The  multiplicity parameter $\nu$ is derived from $\partial F_1(\chi,\nu)
/\partial \nu = 0$ leading to an explicit equation
\begin{widetext}
\begin{equation}\label{eq:Global-n}
\nu = \frac 4{\beta^2J^2} \ \frac {N^{-1}\sum_i \left[\left\langle
\ln\cosh[\beta(h + \eta_i + \lambda J\sqrt{\chi})]\right\rangle_\lambda -
\ln \left\langle \cosh^\nu[\beta(h + \eta_i + \lambda
J\sqrt{\chi})]\right\rangle_\lambda^{1/\nu}\right]} {\chi(2Q + \chi)}\ ,
\end{equation}
\end{widetext}\end{subequations}
 where we denoted $Q \equiv N^{-1}\sum_i M_i^2$.

Global equations~\eqref{eq:Global-equations} complete local stationarity
equations~\eqref{eq:Local-equations}. Free energy,
Eq.~\eqref{:eq:1RSB-TAP},  together with stationarity
equations~\eqref{eq:Local-equations} and~\eqref{eq:Global-equations}
define an analytic theory in the entire space of the input parameters.
They reduce to the TAP theory in the high-temperature phase but
generally differ from it in the spin-glass phase. The spin-glass phase is
characterized apart from local magnetizations also by two global
parameters~$\chi$ and~$\nu$.  The principal difference between free energy
$F_1$ and $F_{ TAP}$ is in the $\lambda$ integral. This integration stands
for thermal equilibration of the replicated spins, that is, for summations
of spin configurations in the phase space determining other TAP
solutions. Alternatively we can understand the $\lambda$-integration
as a thermally weighted averaging of the initial conditions for the
TAP equations. Due to the dependence of  TAP states on the initial
conditions an additive homogeneous internal magnetic field
$\lambda J\sqrt{\chi}$ emerges.
If  the interaction between different TAP solutions (initial and final
configurations of lo\cal magnetizations) vanishes, $\chi =
0$, free energy $F_1$ reduces to $F_{TAP}$.

There are also other situations, when $F_1 = F_{TAP}$. If $\nu=1$,
functional $F_1$ is independent of $\chi$ and we  recover the TAP free
energy. The TAP free energy is recovered also in the limits $\nu\to
\infty$ and $\nu\to 0$. In the former case the $\lambda$-integration
reduces to a saddle point at which $\nu\chi =\Gamma^2 <\infty$. We
explicitly obtain the limiting $\nu\to\infty$ value of free energy
\begin{multline}\label{eq:nInfinity-TAP}
 \bar{F}_1(\Gamma,\bar{\lambda}_i) = - \frac 14\sum_{i,j} \beta J_{ij}^2(1
- M_i^2)(1 - M_j^2) \\  - \frac 12\sum_{i,j} J_{ij}M_i  M_j  + \sum_i M_i
\left[\eta_i + \frac 12 \beta J^2\Gamma^2 M_i \right] \\  + \frac
1\beta\sum_i \left\{ \frac {\bar{\lambda}_i^2}{2} -  \ln \big[2\cosh
[\beta(h + \eta_i +  J\Gamma \bar{\lambda_i}) ]\big]\right\}
\end{multline}
being now a functional of $M_i, \bar{\lambda}_i$ and $\Gamma$.  At the
saddle point $\bar{\lambda}_i = \beta J\Gamma M_i$ and we find that
$\partial \bar{F}_1/\partial \Gamma \equiv 0$, that is, free energy
$F_1$ in the limit $\nu = \infty$ does not depend on $\Gamma$ and we
recover the TAP free energy.

 In the  limit $\nu\to 0$  the annealed randomness in the fluctuating
field $\lambda$ reduces to a quenched one and  the one-level hierarchical
free energy reduces to
%
\begin{multline}\label{eq:nZero-TAP}
 F_1(\chi,0)  =  \frac{\beta J^2N}4 \chi (2 - \chi ) - \frac 14\sum_{i,j}
\beta J_{ij}^2(1 - M_i^2)(1 - M_j^2)\\ - \frac 12\sum_{i,j} J_{ij}M_i  M_j
 + \sum_i M_i \left[\eta_i - \frac 12 \beta J^2\chi M_i \right]\\   - \frac
1\beta\sum_i \int\mathcal{D}\lambda_i   \ln \big[2\cosh[ \beta(h  + \eta_i
+ \lambda_i J\sqrt{\chi}) ]\big] \ .
\end{multline}
In this representation we can absorb the fluctuating field $\lambda_i$
into the internal magnetic field $\eta_i$ and add the Gaussian
$\lambda$-integration to the summation over the lattice sites. After
the substitution $\xi_i = \eta_i + \lambda_i J\sqrt{\chi}$ we find
$\chi = 1 - Q$, where again we denoted $Q = N^{-1}\sum_i M_i^2$, and
recover the TAP free energy.

It is clear from the above analysis that Eq.~\eqref{eq:Global-n} has always
two solutions, one for $\nu < 1$ and the second for $\nu > 1$. In the
former case it is a maximum of  free energy and in the latter one it is a
minimum. We show in the next subsection that the solution for $\nu > 1$ is
an unstable extremum of free energy~\eqref{:eq:1RSB-TAP} and hence the only
physically acceptable, stabilizing  extension of the TAP free energy is
that with $\nu < 1$. Free energy~\eqref{:eq:1RSB-TAP} offers a physical
interpretation of the order parameters  $\chi$ and $\nu$. The last term on
the l.h.s. of Eq.~\eqref{:eq:1RSB-TAP} is the genuine interacting part of
the free nergy. It is  a local free energy due to Ising spins in a random
magnetic field $\lambda_i J\sqrt{\chi}$ due to spin configurations of the
replicated spins (other TAP solutions). The $\lambda$-integral stands for
thermal averaging of the replicated spins and the exponent $\nu < 1$
expresses a  weight with which the replicated spins affect the local
partition function. That is, effectively just $\nu N$ spins are influenced
by configurations of the replicated spins

\subsection{Stability conditions}
\label{sec:Stability}

Saddle-point equations~\eqref{eq:Local-equations}
and~\eqref{eq:Global-equations} should lead to an extremum of free energy
$F_1(\chi,\nu)$. The free energy for fixed homogeneous parameters $\chi$
and $\nu$  as a functional of only local magnetizations $M_i$, when
Eq.~\eqref{eq:Local-eta} for the local magnetic field is used, should be a
minimum. Only then the nonlocal susceptibility is positive semidefinite.
The nonlocal susceptibility in the one-level hierarchical TAP theory is
defined analogously as in the standard TAP theory and reads
\begin{multline}\label{eq:1RSB-nonlocal-chi}
\left(\chi^{-1}\right)_{ij}  = - \beta J_{ij} \\ +
\delta_{ij}\left[\beta^2J^2\left(1 - Q - (1 - \nu)\chi\right) + \frac
1{\chi_{ii}}\right]\ .
\end{multline}
The local inhomogeneous susceptibility in tis case is
 \begin{equation}\label{eq:chi-1RSB}
 \chi_{ii} = 1 - M_i^2 - (1 - \nu)\left[\left\langle
\rho_i^{(\nu)} t_i^2\right\rangle_\lambda -  \left\langle \rho_i^{(\nu)}
t_i\right\rangle^2_\lambda\right]
\end{equation}

The fundamental consistency condition (positivity of the spin-glass
susceptibility) is  Eq.~\eqref{eq:TAP-chi2-stabil} with
the local susceptibility $\chi_{ii}$ from Eq.~\eqref{eq:chi-1RSB} reads
\begin{equation}\label{eq:AT-generalized}
1 \ge \frac {\beta^2J^2}N\sum_i \left[1 - (1 -
\nu)\left\langle\rho_i^{(\nu)} t_i^2\right\rangle_\lambda  - \nu
\left\langle\rho_i^{(\nu)} t_i\right\rangle_\lambda^2\right]^2\ .
\end{equation}
If this condition is fulfilled  free energy $F_1(\chi,\nu)$ from
Eq.~\eqref{:eq:1RSB-TAP} is a physically acceptable and consistent
solution for local magnetizations $M_i$, homogeneous overlap
susceptibility $\chi$ and  multiplicity factors $\nu$. It is evident
from Eq.~\eqref{eq:AT-generalized} that if a TAP solution breaks
condition~\eqref{eq:AT-normal}, that is Eq.~\eqref{eq:AT-generalized} for
$\nu = 1$,  and we  increase $\nu$ to higher values  we worsen the
instability of the TAP solution. To improve upon the incurred instability
of the TAP solution we must evidently decrease the multiplicity factor $\nu$
to values lower than one. That is, we have to maximize  free energy with
respect to the matrix of overlap susceptibilities.
 
If Eq.~\eqref{eq:AT-generalized} does not hold we are unable to find a
stable equilibrium state that would not depend on initial conditions
and would be separable from other macroscopic states by a finite gap
in free energy. The degeneracy of the TAP free energy hence has not
been lifted in free energy~\eqref{:eq:1RSB-TAP} completely. To improve
upon this deficiency we have to go to a theory with a higher number of
hierarchies $K>1$. It is evident that the two-level free energy
$F_2(\chi_1,\nu_1,\chi_2,\nu_2)$ reduces to $F_1(\chi,\nu)$ if either
$\chi_2 =0$ or $\chi_1 = \chi_2$. It is straightforward to demonstrate
that breakdown of condition~\eqref{eq:AT-generalized} leads to an
instability of equality $\chi_2 =0$ and the second overlap
susceptibility $\chi_2$ starts to peel off from its zero value.

In the generalized TAP theory with local magnetizations $M_i$, internal
magnetic fields $\eta_i$ and homogeneous overlap susceptibilities $\chi_1,
\nu_1, \ldots, \chi_K, \nu_K$ as order parameters, minimization of the TAP
free energy w.r.t. local parameters does no longer play an essential role
for stability of macroscopic states. This condition is replaced in the
hierarchical extension of the TAP theory by a more important condition, an
extremum w.r.t. the homogeneous order parameters, overlap susceptibilities
$\chi_l$ with their multiplicities $\nu_l$ for $l=1,2,\ldots, K$. Extremum
of the hierarchical free energy w.r.t. homogeneous parameters leads to an
extremum in thermodynamic inhomogeneity of free energy. Since only $\nu_l
<1$ lead to minimization of  thermodynamic inhomogeneity, we have to
maximize free energy to achieve the least inhomogeneous state.
Free energy $F_1$ may hence also become unstable when the one-level
solution does not maximize free energy and solutions with a higher number
of hierarchical levels (different values for the overlap susceptibilities)
produce a higher free energy.This happens if equation $\chi_2 = \chi_1$
becomes unstable and a new value of $\chi_2 < \chi_1$ emerges. This
happens if the following stability condition is broken~\cite{Janis05a}
\begin{equation}\label{eq:RS-stability}
1 \ge \frac {\beta^2J^2}N\sum_i \left\langle\rho_i^{(\nu)}
(1 - t_i^2)^2\right\rangle_\lambda \ .
\end{equation}
Unlike Eq.~\eqref{eq:AT-generalized} condition~\eqref{eq:RS-stability} gets
stabilized with increasing $\nu$. In the TAP theory with $\chi = 0$ both
conditions coincide.

It is necessary that both conditions, Eq.~\eqref{eq:AT-generalized} and
Eq.~\eqref{eq:RS-stability}, are satisfied for the equilibrium values of
all order parameters so that  free energy~\eqref{:eq:1RSB-TAP}  leads to
stable thermodynamic states for almost all  configurations of spin
couplings. It depends on the equilibrium value of the parameter $\nu$
which of these two conditions is (more) broken and hence responsible for
the eventual instability of the one-level TAP free energy $F_1$. It is
Eq.~\eqref{eq:RS-stability} that makes the solution  for $\nu \to 0$
unstable, TAP free energy  from Eq.~\eqref{eq:nZero-TAP}.  It is
Eq.~\eqref{eq:AT-generalized} that leads to instability of solutions with
$\nu \to 1$, TAP free energy~\eqref{eq:TAP-FE}. Note that in the averaged
theory the relevant instability condition of the extended TAP theory
corresponds to the stability of the one-step RSB scheme.

\subsection{Asymptotic solution near the critical point}
 \label{sec:TAP-Asymptotic}
 
Stationarity equations~\eqref{eq:Local-equations}
and~\eqref{eq:Global-equations} in full generality are difficult  to solve
for a fixed configuration of spin couplings. One can, however, investigate
the behavior of the order parameters close to the spin-glass transition. In
particular, one can explicitly confirm that the TAP solutions become
unstable below the spin-glass transition whenever
condition~\eqref{eq:AT-normal} is broken. We prove in this subsection that
if Eq.~\eqref{eq:AT-normal} is broken the overlap susceptibility $\chi$
becomes positive and the multiplicity factor $\nu\in(0, 1)$ deviates from
its equilibrium value from the high-temperature phase.
 
The small parameter in the low-temperature phase is the overlap
susceptibility. We hence expand all necessary quantities from stationarity
equations~\eqref{eq:Local-equations} into powers of~$\chi$. We will need
the two leading  nontrivial orders.  The asymptotic form of the  local
magnetization  at the AT line reads
\begin{multline}\label{eq:M-m}
M_i \doteq \mu_i - \beta^2J^2(1 - \nu) \mu_i(1 - \mu_i^2)\chi \\
+ \beta^4J^4(1 - \nu)\mu_i(1 - \mu_i^2)\left[ 2 - \nu - (3 -
2\nu)\mu_i^2\right] \chi^2
\end{multline}
 where we denoted $\mu_i = \tanh[\beta(h + \eta_i)]$. In
expansion~\eqref{eq:M-m} we assumed that the internal magnetic field is
fixed, although its stationary value also depends on $\chi$. This
dependence will be evaluated at the end of  our calculations.
 
The difference on the r.h.s. of Eq.~\eqref{eq:Global-chi} must be expanded
into first two orders in $\chi$. We obtain with the above notation
 \begin{multline}\label{eq:chi_ii-asympt}
 \left\langle \rho_i^{\nu)} t_i^2\right\rangle_\lambda -  \left\langle
\rho_i^{(\nu)} t_i\right\rangle_\lambda^2 \doteq \beta^2J^2(1 - \mu_i^2)^2
\chi \\ - \beta^4J^4(1 - \mu_i^2)^2[2 - \nu -(8 - 5\nu)\mu_i^2]\chi^2\ .
 \end{multline}
We will need to expand the global parameter $Q = N^{-1}\sum_i M_i^2$ in
Eq.~\eqref{eq:Global-n}. Also this parameter must be expanded to first two
powers of $\chi$. We obtain directly from  Eq.~\eqref{eq:M-m}
 \begin{multline}\label{eq:q-asympt}
 Q \doteq \left\langle \mu_i^2\right\rangle_{av} - 2 \beta^2J^2(1 -
\nu)\left \langle \mu_i^2(1 - \mu_i^2)\right \rangle_{av} \chi \\ +
\beta^4J^4(1 - \nu) \left\langle \mu_i^2(1 - \mu_i^2)\left[ 5 - 3\nu - (7
- 5\nu) \mu_i^2\right]\right\rangle_{av} \chi^2
 \end{multline}
 where we abbreviated $\langle X_i\rangle_{av} = N^{-1}\sum_i X_i$. This
notation, originating in self-averaging property of local variables, we
also use in the following formulas.
 
Next we denote
 \begin{multline}
 \varphi = \frac 4{\beta^2N}\sum_i \left[\left\langle \ln\cosh[\beta(h +
\eta_i + \lambda J\sqrt{\chi})]\right\rangle_\lambda \right. \\ \left. -
\ln \left\langle \cosh^\nu[\beta(h + \eta_i + \lambda
J\sqrt{\chi})]\right\rangle_\lambda^{1/\nu}\right]\ .\nonumber
\end{multline}
We expand this function to $O(\chi^3)$ and use it together with
Eq.~\eqref{eq:q-asympt} for the evaluation of the expansion of both sides
of Eq.~\eqref{eq:Global-n}. Using the program MATHEMATICA we end up with
\begin{multline}\label{eq:Delta-TAP}
 \Delta = \nu\chi(2Q + \chi) - \varphi \doteq \nu\chi^2\left\{1 -
\beta^2J^2\left\langle(1 - \mu_i^2)^2\right\rangle_{av} \right. \\ \left. +
\frac 23 \beta^4J^4 \chi\left\langle(1 - \mu_i^2)^2\left[3 - 2\nu - (11 -
8\nu)\mu_i^2\right]\right\rangle_{av}\right\}\ .
\end{multline}

Before we proceed with solving the asymptotic forms of
equations~\eqref{eq:Global-chi} and~\eqref{eq:Global-n} we have to
determine the $\chi$-dependence of the equilibrium value of the internal
magnetic field $\eta_i$. It is sufficient for our purposes to expand this
field only to linear power and we replace $\eta_i \to \eta_i^0 +
\chi\dot{\eta}_i$. The local magnetization changes accordingly
\begin{equation}\label{eq:m-expanded}
\mu_i  \doteq m_i + (1 - m_i^2)\chi\beta\dot{\eta}_i
\end{equation}
where we denoted $m_i = \tanh[\beta(h + \eta^0_i)]$  the TAP local
magnetization with  the fluctuating internal magnetic field
$\eta_i^0$  determined by the TAP equation~\eqref{eq:TAP-magf}. We
derive an equation for $\beta\dot{\eta}_i$ from
Eq.~\eqref{eq:Local-eta}. We have
\begin{subequations}
\begin{multline}\label{eq:dot-eta}
\beta\dot{\eta}_i = \beta^2J^2\left[(1 - \nu) + \dot{Q}\right] M_i  \\ +
\sum_j \left[\beta J_{ij} - \delta_{ij} \beta^2J^2(1 - Q)\right]
\dot{M}_j\ .
\end{multline}
Further on, we obtain from Eq.~\eqref{eq:M-m} for $\dot{M}_i =
d M_i/ d\chi$ an asymptotic relation
\begin{equation}\label{eq:dot-m}
\dot{M}_i \doteq (1- m_i^2)[\beta\dot{\eta}_i - \beta^2J^2(1 -
\nu) m_i]\ .
\end{equation}
The equation for $\dot{Q}$ follows directly from
expansion~\eqref{eq:q-asympt}.

Combing the above equations and using the definition for the TAP
susceptibility we come to a solution
\begin{multline}\label{eq:dot-eta-solution}
\beta\dot{\eta}_i \doteq\beta^2J^2(1 - \nu)\bigg[ m_i\\ - 2\beta^2J^2
\frac{\langle m_i^2(1 - m_i^2)\rangle_{av}}{(1 - m_i^2)} \sum_j
\chi^{TAP}_{ij} m_j\bigg] \ .
\end{multline}
\end{subequations}
To reach a representation in closed form we have to evaluate
sums with the linear susceptibility of  type
$N^{-1}\sum_{ij}\chi^{TAP}_{ij}f(m_i)g(m_j)$. We derive an explicit
formula for such sums in  Appendix~\ref{sec:App-linear-susceptibility}.

With explicit expressions for the sums with the nonlocal
susceptibility we have at hand all necessary ingredients to resolve
the asymptotic forms of equations for the global order parameters near
the critical point. We first use Eq.~\eqref{eq:A-chi-sum} to evaluate
\begin{multline}\label{eq:SGsusceptibility-expanded}
\left\langle (1 - \mu_i^2)^2\right\rangle_{av}\\ \doteq  \left\langle (1 -
m_i^2)^2\right\rangle_{av} - 4 \left\langle (1 -
m_i^2)^2m_i\beta\dot{\eta}_i\right\rangle_{av}\chi\\  =  \left\langle (1 -
m_i^2)^2\right\rangle_{av}  + 4\beta^2J^2(1 - \nu) \left\langle m_i^4 (1
- m_i^2)\right\rangle_{av}\chi\\  - 8\beta^4J^4  \left\langle m_i^2 (1
- m_i^2)\right\rangle_{av} \left\langle m_i^2 (1
- m_i^2)^2\right\rangle_{av}\chi
\end{multline}
With this result the asymptotic form of the equation for the overlap
susceptibility reads
\begin{widetext}
\begin{multline}\label{eq:chi-final}
\beta^2J^2 \left\langle(1 - m_i^2)^2\right\rangle_{av} - 1 \doteq
\beta^4J^4\chi \big\{\left\langle(1 - m_i^2)\left[2 - \nu - 2(5 -
3\nu)m_i^2 + (4 - \nu)m_i^4\right]]\right\rangle_{av}\\   +
8 \beta^2J^2(1 - \nu)\left\langle m_i^2(1 - m_i^2)\right\rangle_{av}
\left\langle m_i^2(1 - m_i^2)^2\right\rangle_{av}\big\}
\end{multline}
while the equation for the multiplicity factor $\nu$ can be rewritten
to
\begin{multline}\label{eq:Delta-final}
\beta^2J^2 \left\langle(1 - m_i^2)^2\right\rangle_{av} - 1 \doteq \frac 23
\beta^4J^4\chi \big\{ \left\langle(1 - m_i^2)\left[3 - 2\nu - 2(7 -
5\nu)m_i^2 + (5 - 2\nu)m_i^4\right]]\right\rangle_{av} \\  + 12
\beta^2J^2(1 - \nu)\left\langle m_i^2(1 - m_i^2)\right\rangle_{av}
\left\langle m_i^2(1 - m_i^2)^2\right\rangle_{av}\big\}\ .
\end{multline}
\end{widetext}

Both equations~\eqref{eq:chi-final} and~\eqref{eq:Delta-final} are in
fact defining equations for the overlap susceptibility $\chi$.
Left-hand sides of both equations are identical and become positive in
the low-temperature phase when condition~\eqref{eq:AT-normal} is
broken. Since the solutions from both equations must lead to the same
unique value of $\chi$ we have to equal right-hand sides of these
equations. As a result we obtain an equation for the value of the
parameter $\nu$ along the AT line of critical points. Its solution
reads
\begin{equation}\label{eq:nu-solution}
\nu \doteq \frac {2\langle m_i^2(1 - m_i^2)^2\rangle_{av}}{\langle (1 -
m_i^2)^3\rangle_{av}} \ .
\end{equation}
Parameter $\nu$ obtained from Eq.~\eqref{eq:nu-solution} is the
limiting value of the low-temperature solution at the AT line. It is
positive at finite magnetic field.  This causes no problem, since we know
that the high-temperature solution obeying  the consistency
condition~\eqref{eq:AT-normal} is independent of~$\nu$ (thermodynamically
homogeneous). To determine the deviation of~$\nu$ from its value at the
AT line in the spin-glass phase we had to go to higher orders of the
expansion in~$\chi$.

With the above solution for the multiplicity factor we can use either
Eq.~\eqref{eq:chi-final} or Eq.~\eqref{eq:Delta-final} to determine the
overlap susceptibility $\chi$. The solution for this parameter is physical
only if the r.h.s. of Eqs.~\eqref{eq:chi-final} and~\eqref{eq:Delta-final}
is positive. We can conclude already from Eq.~\eqref{eq:nu-solution} that
this cannot be the case down to zero temperature along the AT line. The
geometric parameter $\nu$ must be smaller than one. We have a critical
value $\nu_c$ of this parameter at which the r.h.s. of
Eqs.~\eqref{eq:chi-final} and~\eqref{eq:Delta-final} vanish, namely
\begin{widetext}
\begin{equation}\label{eq:nu-critical}
\nu_c = 2\ \frac{\langle(1 - m_i^2)(1 - 3 m_i^2)\rangle_{av} \langle(1 -
m_i^2)(1 - 3 m_i^2 + 2 m_i^4)\rangle_{av}} {\langle (1 - m_i^2)(1 - 4
m_i^2)\rangle_{av}^2 - \langle m_i^2(1 - m_i^2)\rangle \langle(1 -
m_i^2)(1 - 2 m_i^2)\rangle_{av} +  \langle m_i^4(1 - m_i^2)\rangle
\langle(1 - m_i^2)(1 - 9 m_i^2)\rangle_{av}}\ .
\end{equation}
\end{widetext}
Using the solution for $\nu$ from Eq.~\eqref{eq:nu-solution} on the l.h.s.
of Eq.~\eqref{eq:nu-critical} we obtain an equation for a critical value
of the magnetic field (temperature) above (below) which the above
asymptotic solution breaks down and we have to go to higher-order terms in
the expansion in the overlap susceptibility. We hence experience a
crossover in the behavior of the homogeneous order parameters along the
instability (AT) line if we go to high magnetic fields. While in low
magnetic fields the overlap susceptibility is determined from a linear
equation~\eqref{eq:chi-final}, we have a quadratic equation determining the
leading asymptotic term near the AT line in high magnetic fields. The
instability of the TAP equation in high magnetic fields
is a rather complex task and will be presented in a separate publication.

\section{Summary and conclusions}
\label{sec:Conclusions}

We analyzed the low-temperature thermodynamics of mean-field models of spin
glasses. In particular, we concentrated on the behavior of thermodynamic
potentials for individual configurations of spin couplings. For this
purpose Thouless, Anderson, and Palmer proposed a construction of a
configurationally-dependent free energy of the Sherrington-Kirkpatrick
model. The derivation of the TAP free energy is, however, valid only if a
convergence or stability condition~\eqref{eq:AT-normal} is obeyed. Typical
configurations of spin couplings in the spin glass phase either do not
allow for solutions of the TAP equations satisfying this condition or
produce a multitude of solutions degenerate in free energy
macroscopically many of which break Eq.~\eqref{eq:AT-normal}. This
situation naturally evokes a number of questions about the TAP
construction: 1) Is it complete? 2) Does it produce stable equilibrium
states? 3) Does the thermodynamic limit exist? Finally, we know that the
exact solution of the SK model is the Parisi RSB scheme. The order
parameters introduced by the replica trick are not manifested in the TAP
thermodynamic potentials. Hence, we should answer another question: 4) At
what stage do the RSB order parameters emerge?

Presently, it is predominantly assumed that the TAP theory is complete as it
is and contains all necessary order parameters from which we can
construct the exact solution. It does not produce a single equilibrium
state, but rather exponentially many locally stable and unstable states
separated by infinite energy barriers and (almost) degenerate in free
energy. Hence a weighted sum~\eqref{eq:TAP-EQ} of local free-energy minima
is to be taken into account to construct a global equilibrium state with
which we can construct the thermodynamic limit. The only information
missing in the TAP thermodynamic potentials is the complexity, i.~e., the
number of available TAP states, local minima of the TAP free energy.  There
is, however,  no trace of the RSB order parameters in the TAP construction
and they are introduced only in course of averaging over the quenched
randomness in spin couplings.
 
In this paper we proposed alternative answers to the above urgent
questions about the TAP construction and its relation to the RSB order
parameters. We explicitly demonstrated that the TAP free energy for
situations with broken stability condition~\eqref{eq:AT-normal} is
unstable The TAP approach becomes incomplete and must be enriched by
new order parameters. The necessity for the enhancement of the TAP
construction emerges due to the need to lift degeneracy in the TAP
free energy that cannot separate stable from unstable states.  Unlike
the existing approaches we do not need to assume impenetrable energy
barriers between different TAP states. We allow for energy flows
between these states if it is thermodynamically convenient and if it leads
to stabilization of equilibrium states. The energy flow between
them is mediated and controlled by new homogeneous order parameters,
overlap susceptibilities.  These additional order parameters are
determined thermodynamically from stationarity equations so that to
achieve a thermodynamically homogeneous free energy with (marginally)
stable equilibrium states. The overlap susceptibilities introduced in
the proposed extension of the TAP construction of a
configurationally-dependent free energy are directly related to the
Parisi RSB order parameters. They coincide after averaging over spin
couplings. Since the configurationally-dependent free energy with
overlap susceptibilities is self-averaging, averaging over randomness
is performed within linear response theory and with the
fluctuation-dissipation theorem as in the case of the SK solution.

We demonstrated in this paper that the TAP construction is incomplete in
the low-temperature phase, the TAP states are unstable and decay into a
composite state described by inhomogeneous local magnetizations and
homogeneous overlap susceptibilities. The extended free energy from which
the physical values of the order parameters are determined is
self-averaging with a well defined equilibrium state and thermodynamic
limit. The RSB order parameters, the overlap susceptibilities,  emerge due
to thermal fluctuations as mediators of  interaction between
different TAP states. Averaging over randomness is harmless and does not
change the structure of the phase space of the order parameters.

When compared with the existing treatments of the thermodynamic behavior of
spin-glass models we can conclude that the hierarchical TAP free
energy~\eqref{eq:RSB-TAP} reduces to the TAP one for equilibrium states
described by local magnetizations satisfying
condition~\eqref{eq:AT-normal}. The proposed extension of the TAP
construction may then seem redundant, since only TAP solutions being local
minima satisfying Eq.~\eqref{eq:AT-normal} are physically relevant. It is,
however, not the case. The proper analytic continuation of the TAP
approach to unstable states guarantees a consistent description of all
states without a tedious way of the separation of locally stable and
unstable solutions of the TAP equations. Moreover, the interaction between
the TAP solutions introduced by the overlap susceptibilities changes the
structure of the underlying phase space and the value of free energy. The
hierarchical TAP theory does not require solving   numerically  the TAP
equations for typical configurations of spin couplings in finite volumes
or to calculate the complexity of the TAP theory. To determine
thermodynamic properties of the SK model we can directly average the
configurationally-dependent free energy in the thermodynamic limit, which
is a significant simplification.

\section*{Acknowledgments}

Research on this problem was carried out within a project AVOZ10100520 of
the Academy of Sciences of the Czech Republic and was supported in part by
Grant No. IAA1010307  of the Grant Agency  of the Academy of Sciences of
the Czech Republic.

\begin{appendix}

\section{Spin glass susceptibility and the resolvent}
\label{sec:App-resolvent}

The averaged local susceptibility $\chi$ and the spin-glass susceptibility
$\chi_{SG}$ can be derived from the resolvent constructed from the inverse
nonlocal susceptibility. The inverse of the nonlocal susceptibility is a
second derivative of free energy and can generally be represented as
\begin{equation}\label{eq:A-inverse-susceptibility}
\left(\chi^{-1}\right)_{ij} = -\beta J_{ij} + \delta_{ij}\left( \frac
1 {\chi_{ii}} + \sum_j\beta^2J^2_{ij}\chi_{jj}\right)\ .
\end{equation}
The resolvent for a complex energy $z$ (scaled by $\beta$ in the same
way as the inverse susceptibility) is defined
\begin{equation}\label{eq:A-resolvent-definition}
G(z) = \frac 1N \text{Tr}\left[z\widehat{1} -
\widehat{\chi}^{-1}\right]^{-1}\ .
\end{equation}
The averaged local susceptibility and the spin-glass susceptibility can be
derived from the resolvent as
\begin{subequations}\label{eq:A-chi-def}
\begin{align}\label{eq:A-chi-loc-def}
\chi = \frac 1N\sum_i \chi_{ii} = - G(0)\, \\ \label{eq:A-chi-sg-def}
\chi_{SG} = \frac 1N\sum_{ij}\chi_{ij}^2 = -\frac{d G(z)}{d z}\bigg|_{z=0}
\end{align}
\end{subequations}
In the Sherrington-Kirkpatrick model  we have
$\sum_j\beta^2J^2_{ij}\chi_{jj} = \beta^2J^2 \chi = -\beta^2J^2 G(0)$. We
now use a theorem of Pastur \cite{Pastur74} for the resolvent of matrices
with off-diagonal elements being Gaussian random variables with variance
$J^2/N$. When applied to the inverse susceptibility we obtain for
$\Delta G(z) = G(z) - G(0)$
\begin{equation}\label{eq:A-resolvent-difference}
\Delta G(z) = - \frac 1N \sum_i \frac
{\chi_{ii}^2(z - \beta^2J^2 \Delta G(z))} {1 - \chi_{ii}(z - \beta^2J^2
\Delta G(z))}\ .
\end{equation}
Using the definition of the spin-glass susceptibility,
Eq.~\eqref{eq:A-chi-sg-def} we obtain
\begin{equation}\label{eq:A-SG-susceptibility}
\chi_{SG} = \frac{\displaystyle\frac 1N\sum_i \frac {\chi_{ii}^2}{(1 +
\beta^2J^2\Delta G(0)\chi_{ii})^2}}{\displaystyle 1 - \frac {\beta^2J^2}N
\sum_i \frac {\chi_{ii}^2}{(1 + \beta^2J^2\Delta G(0)\chi_{ii})^2}}
\end{equation}
Assuming continuity of the resolvent at origin $z=0$ we have $\Delta G(0) =
0$ and we end up with representation~\eqref{eq:TAP-chi2}.

Note that the resolvent representation~\eqref{eq:A-resolvent-difference}
does not exclude a nontrivial solution for $\Delta G(0)$. Setting $z=0$ in
Eq.~\eqref{eq:A-resolvent-difference} we obtain an equation
\begin{equation}\label{eq:A-DeltaG0}
\Delta G(0) = \beta^2J^2\Delta G(0) \frac 1N \sum_i \frac {\chi_{ii}^2}{1 +
\beta^2J^2\Delta G(0)\chi_{ii}}
\end{equation}
allowing for a nontrivial solution if the stability
condition~\eqref{eq:AT-normal} is broken. This nontrivial solution was
used by Plefka in Refs.~\cite{Plefka02a,Plefka02b} in his extension of the
TAP theory. If we choose the nontrivial solution for $\Delta G(0)$ dictated
by analyticity of the resolvent in the complex plane, the spin-glass
susceptibility is no longer  represented by Eq.~\eqref{eq:TAP-chi2} but
rather by Eq.~\eqref{eq:A-SG-susceptibility} and remains positive in the
spin-glass phase. The new parameter $\Delta G(0)>0$ cannot, however, be
derived from a free energy and does not possess a diagrammatic
representation. It is not a proper symmetry-breaking order parameter of a
microscopic origin. Moreover, with this parameter  we break  continuity of
the resolvent and
\begin{equation}\label{eq:A-resolvent-discontinuity}
\lim_{z\to0}\ G(z) \neq G(0) = -\frac 1N\sum_i \chi_{ii}\ .
\end{equation}
The last equality is the definition of the averaged local susceptibility,
Eq.~\eqref{eq:A-chi-loc-def}. The discontinuity  makes a physical
interpretation and explanation of the order  parameter $\Delta G(0)$
difficult.  We can only observe that positivity of $\Delta G(0)$  formally
expresses a deviation from the fluctuation-dissipation theorem. There is no
evidence or indication that the TAP solutions really lead to a
discontinuous resolvent and $\Delta G(0) > 0$ in the spin-glass phase. An
alternative way how to reach thermodynamic consistency and positivity of
the spin-glass susceptibility within a microscopic construction provided
by the hierarchical free energy with a fluctuation-dissipation theorem in
the extended phase space with real spin replicas  is offered in this
paper.

\section{Sums with the nonlocal mean-field susceptibility}
\label{sec:App-linear-susceptibility}

The mean-field approximation is a single-site theory in that it
effectively decouples distinct lattice sites. The decoupling of
distinct lattice sites leads to a simplification of sums with nonlocal
functions. These sums can be converted in the mean-field theory to
uncorrelated lattice sums  with site-local functions.
Correlation between different sites enters mean-field expressions only via
homogeneous global parameters being again uncorrelated sums over lattice
sites.
 
In the spin-glass mean-field theory we are interested in sums with the
nonlocal susceptibility of form
\begin{equation}\label{eq:A-sum-definition}
C[f,g] = \frac 1N \sum_{ij}\chi_{ij}f(m_i)g(m_j)
\end{equation}
The only nonlocal term in the susceptibility is the spin exchange
$\beta J_{ij}$.  It is the off-diagonal part of the susceptibility
that makes the evaluation of sums from Eq.~\eqref{eq:A-sum-definition}
difficult. We hence use the following representation for the nonlocal
susceptibility
\begin{align}\label{eq:A-chi-nonolocal}
\chi_{ij} &= \chi_{ii}\left[\delta_{ij} + \sum_k\!{}^\prime \beta
J_{ik}\chi_{kj}\right]\nonumber\\ & = \chi_{ii} + \chi_{ii}\left[\beta
J_{ij} + \sum_k \beta J_{ik}\chi_{kk}\beta J_{kj}\right. \nonumber \\
&\left. + \sum_{k\neq j}\sum_{l\neq i} \beta J_{ik}\chi_{kk}\beta
J_{kl}\chi_{ll} \beta J_{lj} +\ldots \right]\chi_{jj}
\end{align}
where the primed sum does not allow for repetition of site indices. It
means that only self-avoiding random walks contribute to the inverse
matrix in the formal solution to Eq.~\eqref{eq:A-chi-nonolocal}.

Representation~\eqref{eq:A-chi-nonolocal} can easily be proved by a
diagrammatic expansion when the definition of the TAP
susceptibility~\eqref{eq:TAP-isuscept} is used. We  successively
exclude repeating site indices in the multiple sums of the expansion for
the inverse of the r.h.s. of expression~\eqref{eq:TAP-isuscept}. The
diagonal element of the susceptibility $\chi_{ii}$ was determined along
this line e.~g. in Ref.~\cite{Bray79}.

Since the site indices in Eq.~\eqref{eq:A-chi-nonolocal} are decoupled we
can use the following functional representation for the spin exchange of
the SK model
\begin{equation}\label{eq:A-J-functional}
\beta J_{ij} = \frac{\beta^2J^2}N\ \left[ \nabla_i m_j + m_i \nabla_j
\right]\ .
\end{equation}
We denoted $\nabla_i \equiv \chi_{ii}\partial/\partial m_i$.
Representation~\eqref{eq:A-J-functional} is a consequence of the fact that
just squares of the spin coupling $J_{ij}$ contribute to the sum $C[f,g]$.
  The paired spin exchange to the given one $J_{ij}$ connecting lattice
sites $i$ and $j$ can be extracted from the end-point functions of local
magnetizations $m_i$ and/or $m_j$.  A more detailed proof of
Eq.~\eqref{eq:A-J-functional} can be found in Ref.~\cite{Janis05b}.
 
 Using Eq.~\eqref{eq:A-J-functional} we can represent the off-diagonal
susceptibility $\widetilde{\chi}_{ij} = \chi_{ij} - \chi_{ii}\delta_{ij}$
as
\begin{multline}\label{eq:A-chi-off}
\widetilde{\chi}_{ij} = \frac{\beta^2J^2}N \left\{ \nabla_i \chi_{ii} m_j
\chi_{jj} + m_i \chi_{ii}\nabla_i \chi_{jj} \right. \\ \left. +  \nabla_i
X_j + m_i \chi_{ii} Y_j\right\}
\end{multline}
where we denoted global parameters $X_j = \sum_k m_k \widetilde{\chi}_{kj}$
 and $Y_j = \sum_k \nabla_k \widetilde{\chi}_{kj}$. Note that the
differential operator $\nabla_i$  acts to the right on functions of the
local magnetization $m_i$ only. The lattice sums in the definition of the
global parameters $X_j$ and $Y_j$ should avoid the fixed index~$j$. In the
mean-field approximation we can neglect this restriction, since the
difference is only of order $O(N^{-1})$.

It is straightforward to find from Eq.~\eqref{eq:A-chi-off}  an equation
for
\begin{multline}\label{eq:A-X-eq}
X_i = \beta^2J^2 \left\{ \langle\nabla_k
m_k\chi_{kk}\rangle_{av}m_i\chi_{ii} + \langle m_k^2\chi_{kk}\rangle_{av}
\nabla_i\chi_{ii}\right. \\ \left. + \langle\nabla_k
m_k\chi_{kk}\rangle_{av} X_i + \langle m_k^2 \chi_{kk}\rangle_{av}
Y_i\right\}
\end{multline}
where we denoted as in the main text $\langle X_k\rangle_{av} \equiv
N^{-1}\sum_k X_k$. Analogously we find
\begin{multline}\label{eq:A-Y-eq}
Y_i = \beta^2J^2 \left\{ \langle\nabla_k\nabla_k
\chi_{kk}\rangle_{av}m_i\chi_{ii} + \langle
\nabla_k m_k\chi_{kk}\rangle_{av} \nabla_i\chi_{ii}\right. \\ \left. +
\langle\nabla_k \nabla_k\chi_{kk}\rangle_{av} X_i + \langle \nabla_k m_k
\chi_{kk}\rangle_{av} Y_i\right\} \ .
\end{multline}

To represent the solution for these parameters concisely  we  denote $l =
\beta^2J^2\langle(1 - m_i^2)^2\rangle_{av}$ and $r = \beta^2J^2\langle
m_i^2(1 - m_i^2)\rangle_{av}$. Then
\begin{equation}\label{eq:A-X-solution}
X_i = \frac{(1 - l)(l - 2 r) m_i\chi_{ii} + r \nabla_i\chi_{ii}} {(1 - l)^2
+ 2 r(2 - l)}
\end{equation}
and
\begin{equation}\label{eq:A-Y-solution}
Y_i = (2 l - r)\ \frac{- 2 m_i\chi_{ii} +  (1 - l) \nabla_i\chi_{ii}} {(1 -
l)^2 + 2 r(2 - l)}\ .
\end{equation}

Inserting Eqs,~\eqref{eq:A-X-solution} and~\eqref{eq:A-Y-solution} in
Eq.~\eqref{eq:A-chi-off} we obtain
\begin{multline}\label{eq:A-chi-solution}
\chi_{ij} = \chi_{ii} \delta_{ij} + \frac{\beta^2J^2}{N\left[(1 - l)^2 + 2
r(2 - l)\right]} \\  \times\left\{ (1 + 2 r - l)  \left[
\nabla_i\chi_{ii} m_j \chi_{jj} + m_i \chi_{ii} \nabla_j\chi_{jj} \right]
\right. \\ \left. - 2 (l - 2 r) m_i \chi_{ii} m_j \chi_{jj} + r \nabla_i
\chi_{ii} \nabla_j \chi_{jj}\right\}\ .
\end{multline}
Equation~\eqref{eq:A-chi-solution} hold only in the leading $N^{-1}$
order. Hence the second term on the r.h.s. contributes only to the
off-diagonal part and to lattice sums with the nonlocal susceptibility.

This representation is still a rather complicated expression.
Fortunately, we need to know  for our purposes the nonlocal susceptibility
only along the AT line for which $l=1$. In this case the nonlocal
susceptibility reduces to
\begin{multline}\label{eq:A-chi-ATline}
\chi_{ij} = \chi_{ii} \delta_{ij} \\  + \frac{\beta^2J^2}{2N} \ \left[ 2
\nabla_i\chi_{ii} m_j \chi_{jj} + 2 m_i \chi_{ii} \nabla_j\chi_{jj} +
 \nabla_i \chi_{ii} \nabla_j \chi_{jj} \right] \\  -
\frac{\langle(1 - m_k^2(1 - 3 m_k^2)\rangle_{av}}{\langle m_k^2(1 -
m_k^2)\rangle_{av}\langle (1 - m_k^2)^2\rangle_{av}} \ m_i \chi_{ii} m_j
\chi_{jj}\ .
\end{multline}
Using this result for functions $f(m_i) = m_i(1 - m_i^2)$ and $g(m_j) =
m_j$  in Eq.~\eqref{eq:A-sum-definition} we  find an explicit
representation for a sum with the nonlocal susceptibility at the AT line
needed in Eq.~\eqref{eq:SGsusceptibility-expanded}
\begin{widetext}
\begin{equation}\label{eq:A-chi-sum}
\frac 1N \sum_{ij} \chi_{ij}m_i(1 - m_i^2)m_j = \frac {\beta^2J^2}2 \
\left\langle(1 - m_k^2)^2\right\rangle_{av} \left\langle(1 -
m_k^2)(1 - 3 m_k^2) \right\rangle_{av}\ .
\end{equation}
\end{widetext}
Notice that the nonlocal susceptibility~\eqref{eq:A-chi-solution} diverges
at the critical point of the SK model only at zero magnetic field where
$r=0$.
\end{appendix}

\end{document}